# Thermodynamic properties in the normal and superconducting states of $Na_xCoO_2 \cdot yH_2O$ powder measured by heat capacity experiments

B. Lorenz[1], J. Cmaidalka[1], R. L. Meng[1], and C. W. Chu[1,2,3]

[1]Department of Physics and TCSAM, University of Houston, Houston, TX 77204-5002

[2]Lawrence Berkeley National Laboratory, 1 Cyclotron Road, Berkeley, CA 94702

[3]Hong Kong University of Science and Technology, Hong Kong, China

**Abstract**

The heat capacity of superconducting $Na_xCoO_2 \cdot yH_2O$ was measured and the data are discussed based on two different models: The BCS theory and a model including the effects of line nodes in the superconducting gap function. The electronic heat capacity is separated from the lattice contribution in a thermodynamically consistent way maintaining the entropy balance of superconducting and normal states at the critical temperature. It is shown that for a fully gapped superconductor the data can only be explained by a reduced ($\approx$ 50 %) superconducting volume fraction. The data are compatible with 100 % superconductivity in the case where line nodes are present in the superconducting gap function.

## 1. Introduction

Superconductivity in two-dimensional $CoO_2$-layers was recently discovered in the layered cobalt oxyhydrate $Na_xCoO_2 \cdot yH_2O$ [1]. The possible similarity to the high-$T_c$ superconductivity in $CuO_2$-planes of the cuprates raised a tremendous interest in this compound despite its rather low transition temperature of $T_c < 5$ K. The compound seems to be the first layered oxide involving 3d-transition metals other than copper. The intercalation of water molecules results in an expansion of the c-axis and enhances the two-dimensional character of the structure. Electron doping is achieved by reducing the sodium content to about 0.3 to 0.35. The magnetic susceptibility exhibits a diamagnetic drop below $T_c$ and the high field behavior resembles that of the high-$T_c$ copper oxides [1,2]. An unconventional nature of superconductivity in this compound was suggested based on the observed upper and lower critical magnetic field parameters [2]. The phase diagram of $Na_xCoO_2 \cdot yH_2O$ was studied as a function of the Na content x (doping) and a maximum of $T_c$ was found similar to the phase diagram of the cuprate superconductors [3]. Recent high-pressure measurements indicate a decrease of $T_c$ with hydrostatic pressure where the pressure coefficient of $T_c$ is compatible with those of electron-doped copper oxide superconductors [4]. Based on the first experimental data various models for the superconductivity and the pairing symmetry have been proposed [5,6,7,8,9,10].

One of the questions raised by experimentalists and theoreticians is the symmetry of the order parameter or the superconducting gap. The results of $^{59}$Co NMR and NQR experiments are contradictory. Some investigations suggest the existence of a fully gapped superconducting state [11,12] but most recent results are in favor of the existence of line nodes in the gap function [13]. On the other hand, heat capacity measurements at low temperature have been interpreted in terms of non s-wave symmetry of the order parameter and the existence of point nodes in the superconducting gap function was proposed [14]. However, as will be discussed later, the data presented in Ref. 14 violate an important thermodynamic requirement, namely the entropy balance of normal and superconducting states at $T_c$ and the conclusions are therefore questionable. In a different report, a hump in the specific heat observed at 6 K was also ascribed to the onset of superconductivity in $Na_xCoO_2 \cdot yH_2O$ although the $T_c$ estimated by magnetic and resistive measurements appeared to be clearly lower (4.5 to 5 K) [15].

Magnetic measurements are frequently used to determine the superconducting volume fraction but for powders of small particles the effect of penetrating magnetic fields into the grains may reduce the diamagnetic signal and make it very difficult to extract the correct superconducting volume. As shown in SEM images of the hydrated powder of $Na_xCoO_2 \cdot yH_2O$ the grain size is rather small (1 to 20 µm) and the larger particles split into slabs of typical width less than 1 µm due to the intercalation of $H_2O$ molecules [15,16]. In fact, most of the published data on dc magnetization measurements show a low-temperature diamagnetic response between 10 and 30 % of the maximum value of $-1/4\pi$ [1,2,11,14,15,17]. Nearly 100 % field screening was only claimed for samples synthesized at Princeton University [3,18]. However, the dc-magnetization vs. field data shown in the inset of Fig. 2 of Ref. 3 do not support the authors conclusion of 100 % shielding but are more consistent with a reduced shielding signal of about 20 to 25 %. The large differences in the magnetization data and the derived shielding fractions reported by different groups is a puzzle in view of the fact that x-ray diffraction measurements do not indicate the presence of a considerable amount of second phases for most of the published data on $Na_xCoO_2 \cdot yH_2O$. Even for "optimally doped" samples with $T_c$ as high as 5 K the superconducting shielding signal can be as low as 15 %. The question arises if the magnetic susceptibility data can be used to get a reliable estimate of the superconducting volume fraction in the compound. Alternative experiments measuring thermodynamic quantities such as specific heat are, therefore, of particular interest and should be compared with the magnetic data. The advantage of specific heat measurements is the volume character of the quantity that allows us to distinguish between minor superconducting phases (easily picked up by resistivity measurements) and bulk superconductivity.

Another obstacle for many experiments on $Na_xCoO_2 \cdot yH_2O$ is the form of the samples. Most measurements have been conducted with powdered samples since the sintering of polycrystalline bulk pellets at higher temperature is not possible (the powder, if heated above room temperature, loses water very rapidly and turns into a non-superconducting structure [16,18]). The compression of powder at room temperature produces extremely porous pellets and poor inter-grain coupling so that the resistivity below $T_c$ does not reach the expected zero value [1,14]. The effect of cold compression on superconducting $Na_xCoO_2 \cdot yH_2O$ powder was shown to result in a systematic decrease of $T_c$ and the diamagnetic signal at 2 K with the force of compression [19]. Since it is not clear whether this effect is due to the weak link nature of the specimen (as suggested in Ref. 1) or if it is related to a reduction of the superconducting volume it is necessary to perform the heat capacity measurements using as synthesized powder of $Na_xCoO_2 \cdot yH_2O$. Heat capacity measurements of powdered samples can be conducted by mixing the specimen with a low-temperature grease to establish good thermal connectivity to the heater. If the heat capacity of the grease is well known it can be subtracted from the raw data.

## 2. Experimental

The $Na_xCoO_2 \cdot yH_2O$ powder was synthesized as described previously [1,4]. The x-ray spectrum shows that the powder is single phase with the hexagonal space group $P6_3/mmc$ (a=2.820 Å, c=19.59 Å). The onset of the superconducting transition at 4.7 K was determined by magnetic susceptibility measurements. The diamagnetic shielding signal at 2 K (zero field cooling) of about 16 % of $-1/(4\pi)$ is comparable with data of Ref. [1]. For the heat capacity measurements the powder was uniformly mixed with the Wakefield Thermal Joint Compound, Type 120 (Wakefield Engineering). The specific heat of this grease is well known [20] and was also measured in a separate run. The experiments were performed in high vacuum using the Physical Property Measurement System (PPMS, Quantum Design). The heat capacity of the thermal grease was subtracted from the raw data according to the mixing ratio.

## 3. Results and Discussion

Figure 1 (upper left inset) shows the sample heat capacity, $C_p/T$, as function of T in the range between 10 K and 2.2 K. The peak close to 4.5 K clearly indicates the superconducting transition. The thermodynamic $T_c$=4.77 K as estimated from an entropy conserving construction is in good agreement with the estimate from the magnetization data. To extract the electronic specific heat from the data the lattice contribution needs to be subtracted. The lattice specific heat is usually estimated by plotting $C_p/T$ versus $T^2$. Within the Debye theory the lattice heat capacity at low temperatures can be expanded into a power series of T starting with the third order term, $\beta T^3$. The electronic contribution from free carriers in the normal state is proportional to T. Therefore, the $C_p/T$ vs. $T^2$ plot is expected to be linear (if only the lowest order of the expansion contributes to the lattice heat capacity) and the slope $\beta$ defines the low temperature contribution of the lattice to $C_p(T)$. The extrapolation to T=0 yields the electronic part (Sommerfeld constant, $\gamma$). Figure 1 (main panel) shows the $C_p/T$ vs. $T^2$ plot for the current data. A simple linear extrapolation of the normal state data just above $T_c$ yields the parameters $\beta$=0.45 mJ/mol K$^4$ and $\gamma$=15.7 mJ/mol K$^2$. Both parameters are in agreement with the results given in Ref. 14 and 15. However, a closer inspection of the electronic heat capacity after subtraction of the lattice contribution $\beta T^3$ reveals a thermodynamic inconsistency of the result. The laws of thermodynamics require that the entropies of normal and superconducting states are equal at $T_c$ since the superconducting transition is a second order phase transition. The entropy is calculated by integrating $C_p/T$:

$$S(T) = \int_0^T \frac{C_p(T')}{T'} dT' \qquad (1)$$

The thermodynamic demands can be visualized in a simple way by plotting the electronic part of $C_p/T$ as function of temperature (lower right inset in Fig. 1). The dashed line is the base line and corresponds to the value of $\gamma$. The requirement that the entropies of both, superconducting and normal states, be equal at $T_c$ has a geometric interpretation: The areas between the superconducting $C_e/T$ above and below the base line have to be equal. This condition is obviously violated in the construction of Figure 1 (inset). Although the heat capacity was measured only to about 50 % of $T_c$ (because of experimental limitations) we may use the dotted line in the inset of the figure as an upper limit for $C_e(T)/T$ at low temperature. An integration of the two areas yields the values 6.1 mJ/mol K and 13.8 mJ/mol K, respectively. The large difference of these two integrals is an indication that the linear extrapolation of the lattice contribution from data just above $T_c$ (Fig. 1) is not justified and non-linear terms have to be taken into account. It should be noted that similar linear extrapolations applied in Ref. 14 and 15 obviously lead to thermodynamic inconsistencies similar or even worse to those discussed above (see for example Fig. 3 of Ref. 14 and Fig. 3 of Ref. 15). Therefore, some of the conclusions drawn in the previous publications may not be valid.

In order to establish thermodynamic consistency higher powers in the expansion of the lattice heat capacity have to be taken into account. The total heat capacity in the normal state is written as

$$C_n(T) = \gamma T + \beta T^3 + AT^5 \quad . \tag{2}$$

The linear first term represents the contribution of the electronic system with the Sommerfeld constant $\gamma$ and the following two terms are the lattice specific heat taking into account the two lowest order (up to $T^5$) contributions. The coefficient A will be treated as an additional fit parameter. Note that in this respect the expansion (2) deviates from the strict Debye model in which A is not an independent parameter but it is correlated with $\beta$. However, the Debye model by itself is an approximation to the lattice heat capacity of real solids and deviations are frequently observed and expressed by a temperature dependent Debye temperature. The coefficients of equation (2) have to be determined under the condition that the entropy balance between normal and superconducting states is preserved. One major uncertainty in the previous and current experiments is the low temperature behavior of the superconducting specific heat. It is interesting to note that the data below $T_c$ shown in the inset of Fig. 1 suggest a power law with the exponent 3, $C_p(T) \sim T^3$, ($T<T_c$). A similar behavior was observed and attributed to the existence of point nodes in the superconducting gap function [14]. However, the differences in the analytic temperature dependence of the heat capacity due to various gap functions (gapless, point and line nodes, fully gapped) become essential only at very low temperature [21] and the current "high"-temperature data (i.e. $T>T_c/2$) cannot decide whether there exist nodes in the gap function.

In order to improve the extrapolation of the normal state $C_p$ and to extract the lattice contribution in a thermodynamically consistent way we also need to know the low temperature heat capacity in the superconducting state. Since our data are limited to $T>T_c/2$ we extrapolate to low temperatures based on different models: (i) Assuming a fully gapped superconductor we use the BCS data for the heat capacity (tabulated by Mühlschlegel [22]). (ii) We discuss the consequences of the presence of line nodes in the superconducting gap function.

### 3.1    Extrapolation according to the BCS theory

The specific heat data in the BCS model have been tabulated in Ref. 22. The lattice contribution is fitted to the experimental data by taking into account higher order terms in the low temperature expansion of the specific heat (equation 2) and requiring thermodynamic consistency, i.e. the entropy balance between superconducting and normal states. It turns out that a fit to the experimental data can only be achieved by assuming that the superconducting volume fraction is well below 100 % of a perfect superconductor. Using the tabulated BCS data we get a reasonable fit to the measured heat capacity with a superconducting volume fraction of about 50 % of the total sample volume. The Sommerfeld constant $\gamma=12.5$ mJ/mol K$^2$ is in good agreement with previously reported values [14,15]. The remaining fit parameters ($\beta$, A) are listed in Table 1. The data and the fit functions in normal and superconducting states are shown in

Fig. 2. The peak of $C_e(T)$ is very sharp at the superconducting transition and the thermodynamic requirement of entropy conservation is fulfilled.

The BCS model for the superconducting heat capacity applies to conventional superconductors with a gap opening across the whole Fermi surface (no nodes). The fits have been conducted in a thermodynamically consistent (i.e. entropy preserving) way. The major consequence of the procedure is a far less than 100 % superconducting volume. However, the value of about 50 % estimated from the heat capacity data is considerably higher than the 16 % of magnetic shielding signal obtained for the present sample from the low-temperature (zero field cooling) magnetization measurements and the 15 to 25 % deducted from magnetization measurements in most of the previous reports [1,2,11,14,15,17]. This clearly indicates that the magnetic shielding measurements are not suitable to determine the superconducting volume fraction, probably due to the weak link nature of the samples as suggested in Ref. 1. In contrast, the measured heat capacity is a thermodynamic volume quantity that is not hampered by weak link or field penetration effects.

The 50 % estimated superconducting volume fraction appears surprising in view of the fact that x-ray spectra of our $Na_xCoO_2 \cdot yH_2O$ powder did not show any indication of second or impurity phases. All lines are very sharp and can be indexed within the $P6_3/mmc$ structure. One possible explanation could be a small variation of the water or sodium content throughout the specimen since it was shown that superconductivity in $Na_xCoO_2 \cdot yH_2O$ exists only in a narrow range of x close to 0.3 [3]. However, it is essential to understand that the conclusion of ~50 % superconducting volume was derived under the assumption that the superconducting gap function has no nodes across the Fermi surface. This assumption needs to be reconsidered since recent $^{59}$Co NQR experiments suggested the existence of line nodes in the gap function [13]. In the following section we extrapolate our heat capacity data to low temperatures using a model for the gap function with line nodes.

### 3.2 Line node model for the superconducting gap function

The presence of line nodes in the gap changes the low temperature behavior of the electronic specific heat to a power law with exponent 2 [21]:

$$C_e(T) = BT^2 , \qquad (T \ll T_c) \qquad . \qquad (3)$$

The best fit of equations (2) (T > T$_c$) and (3) (T < T$_c$) to the experimental data is shown in Fig. 3. The fit preserves the entropy balance and results in a 100 % superconducting volume fraction. The fit parameters are listed in Table 1.

It is remarkable that the line node model is the only theory that explains the experimental C$_p$(T) data in a thermodynamically consistent way and with 100 % of superconducting volume fraction. Other characteristic quantities can be derived from the model and the data. The relative jump of the electronic heat capacity at T$_c$, $\delta C_e(T_c)/\gamma T_c$ = 0.99, is clearly smaller than the corresponding BCS-value of 1.43. This small jump of C$_e$(T$_c$) is unusual and could be related to the line nodes property of the gap function. A smaller (than BCS) value for $\delta C_e(T_c)/\gamma T_c$ has also been reported for MgB$_2$ and it was shown that it is a consequence of the existence of two superconducting gaps with different magnitude at the Fermi surface. In Na$_x$CoO$_2 \cdot y$H$_2$O there is no indication of a similar two-gap scenario.

Unfortunately, the present data acquired in the experimentally limited temperature range of $T > T_c/2$ cannot uniquely distinguish between the different models discussed in this section. In order to decide whether the superconducting volume fraction is 100 % or less and to derive definite conclusions about the superconducting gap structure the heat capacity measurements have to be extended into the very low temperature range ($T \ll T_c$). Another uncertainty is the lattice contribution to the specific heat that has to be subtracted to derive the superconducting heat capacity. A more precise estimate of the lattice contribution, for example by suppressing the superconducting state in very high magnetic fields, is desirable and will enhance the accuracy of the derived electronic term, $C_e(T)$, in particular in the superconducting state. In any case it is important to check the resulting electronic contribution to the specific heat with respect to its thermodynamic consistency. The entropies of the normal and superconducting states must be equal at $T_c$. If this balance is not fulfilled the conclusions derived from the heat capacity data are questionable.

## 4. Summary and Conclusions

We have measured the heat capacity of a $Na_xCoO_2 \cdot yH_2O$ powder above 2 K in the normal and superconducting states. The lattice contribution to the specific heat was subtracted in a thermodynamically consistent way maintaining the entropy balance of both states at $T_c$. Different scenarios for the superconducting gap function are discussed and compared with the experimental data. For a fully gapped superconductor the data can only be explained by assuming that the superconducting volume fraction is about 50 % of the total volume. The origin of the non-superconducting phase (or the 50 % of free carriers not participating in the pairing) is not clear. Our data are compatible with 100 % superconductivity only in the case where line nodes are present in the superconducting gap function. Heat capacity measurements at very low temperature are of particular interest since they can distinguish between the different models.

**Acknowledgments**

This work is supported in part by NSF Grant No. DMR-9804325, the T.L.L. Temple Foundation, the J. J. and R. Moores Endowment, and the State of Texas through the TCSAM and at LBNL by the Director, Office of Energy Research, Office of Basic Energy Sciences, Division of Materials Sciences of the U.S. Department of Energy under Contract No. DE-AC03-76SF00098.

Table 1: Characteristic parameters of the heat capacity in normal and superconducting states as obtained for the two models discussed in Section 3.

|  | BCS theory | Gap function with line nodes |
|---|---|---|
| $\gamma$ [mJ/mol K$^2$] | 12.5 | 10.8 |
| $\beta$ [mJ/mol K$^4$] | 0.56 | 0.62 |
| A [mJ/mol K$^6$] | -0.0012 | -0.0015 |
| T$_c$ [K] | 4.77 | 4.77 |
| B [mJ/mol K$^3$] |  | 4.5 |
| Sc. volume fraction | 50 % | 100 % |
| $\delta C_e(T_c) / \gamma T_c$ | 1.43 | 0.99 |

Figure Captions:

Fig. 1. $C_p/T$ plotted as function of $T^2$. The upper left inset shows the raw data. The lower right inset shows the resulting electronic heat capacity after subtracting a lattice contribution proportional to $T^3$ and neglecting higher order terms. Obviously, the entropy balance between normal and superconducting states is violated (The numbers in the lower inset measure the area between the normal and superconducting parts of the curves. Thermodynamics requires that these numbers are equal).

Fig. 2. Fit of the BCS-values for the heat capacity (full line) to the experimental data (open circles). The extrapolation of the normal state $C_p$ below $T_c$ is shown as the dashed line. The superconducting volume fraction was estimated as 50 %. The inset shows the electronic part, $C_e(T)/T$. Note that the entropy balance is preserved.

Fig. 3. Fit of a model based on a gap function with line nodes to the experimental heat capacity. The full line indicates the fit function in normal and superconducting states. The dashed line is the extrapolation of the normal state $C_p$. The superconducting volume fraction is 100 %. The entropy balance (inset) is preserved.

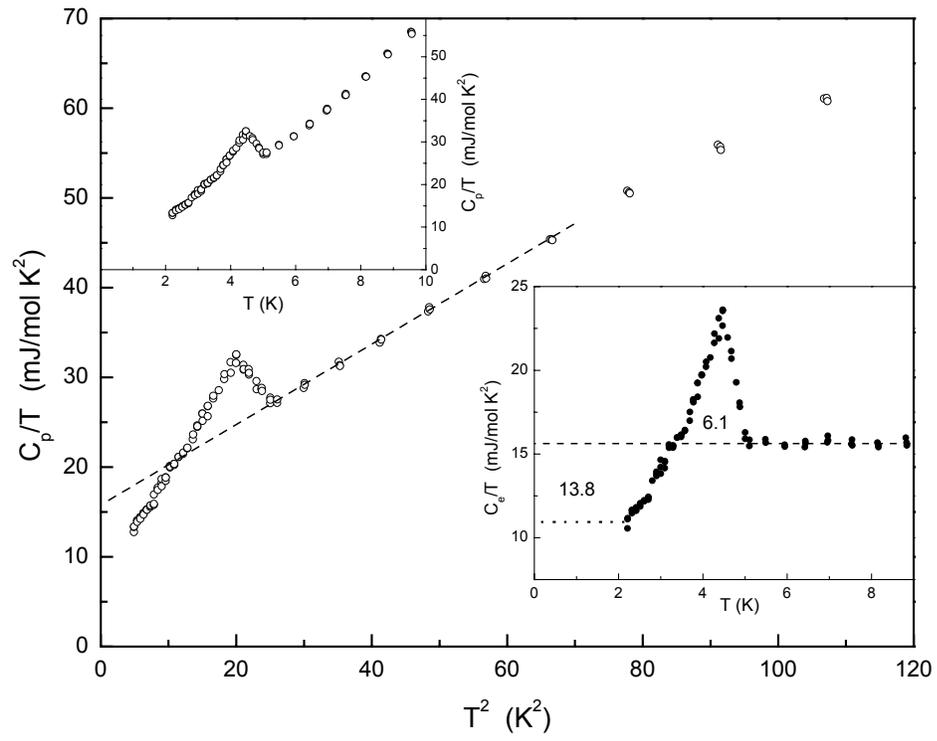

Fig. 1.

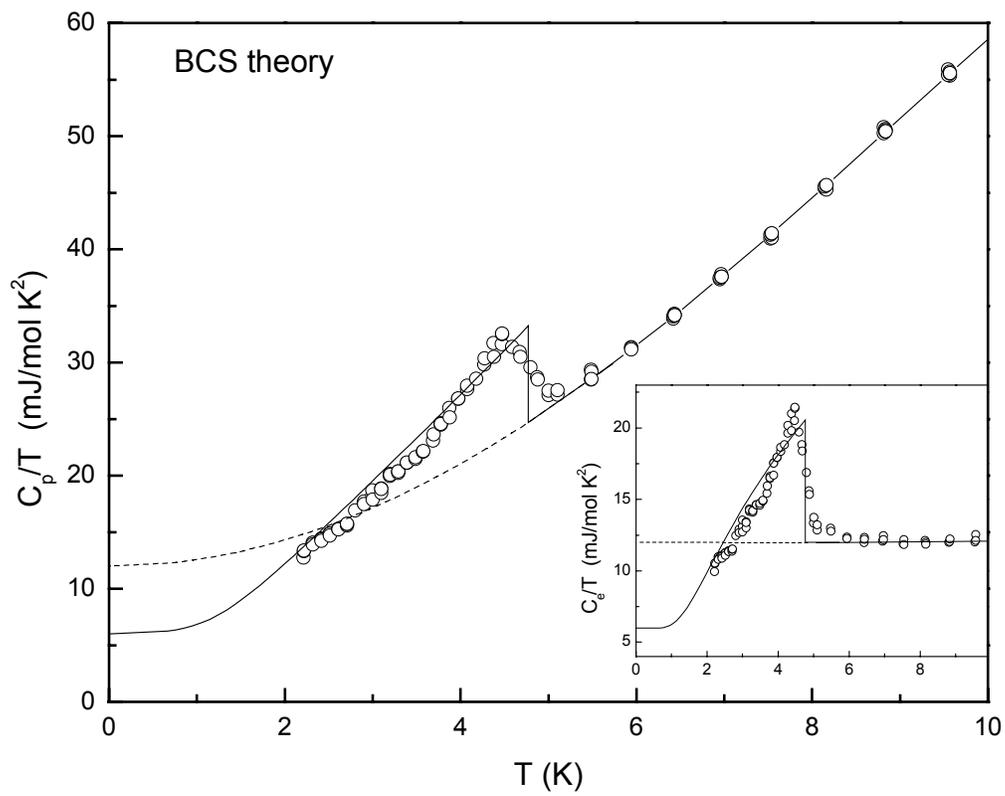

Fig. 2.

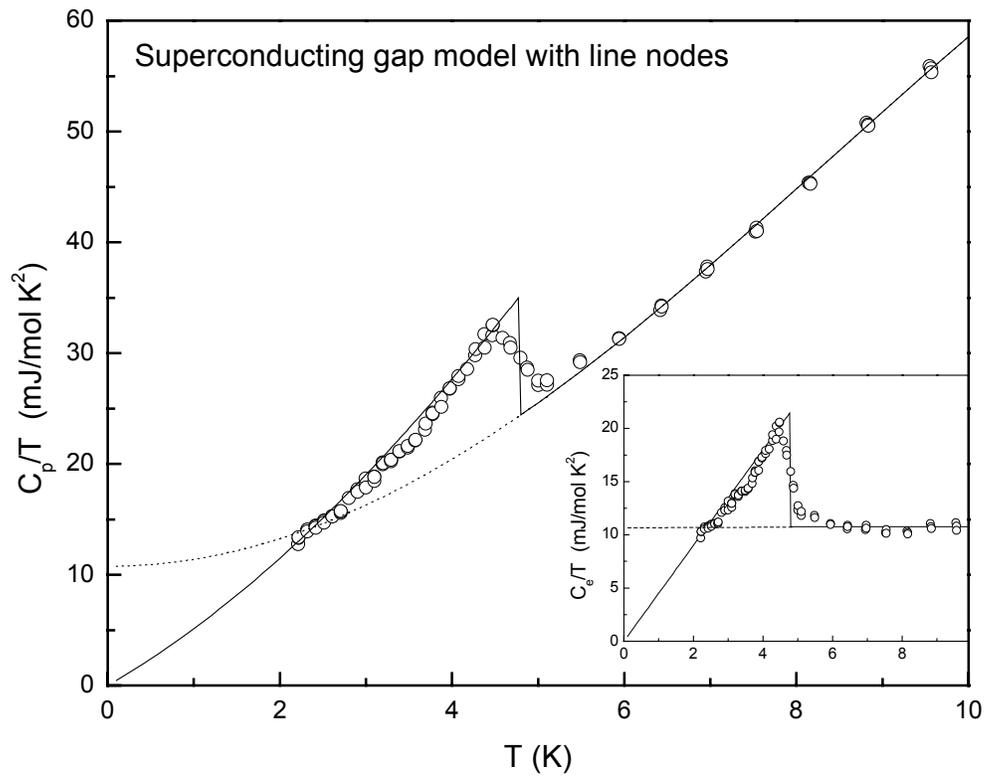

Fig. 3.